# The upgrade of the RPC-based ALICE Muon Trigger

**A. Ferretti,**[a][*] **on behalf of the ALICE collaboration**

[a] *INFN and University of Torino, Via Pietro Giuria 1, 10125, Torino, Italy*

E-mail: alessandro.ferretti@unito.it

ABSTRACT: The ALICE Muon Trigger is currently yielded by a detector currently composed of 72 Bakelite single-gap Resistive Plate Chambers operated in maxi-avalanche mode, arranged in four 5.5x6.5 $m^2$ detection planes. In order to meet the requirements posed by the forthcoming LHC high luminosity runs starting from 2021 onwards, in which ALICE will be read out in continuous mode, the Muon Trigger will become a Muon Identifier and will undergo a major upgrade. In the current setup, signals from about 21k strips are discriminated by 2400 non-amplified Front End (FEE) cards, whose thresholds are provided by external analog voltages (one for each chamber side). All these cards will be replaced with discriminators equipped with a pre-amplification stage which will allow a reduction in the operating high voltage of the detectors, thus prolonging their lifetime. Furthermore, their reference thresholds will be passed via wireless (and I2C chained per chamber side) allowing the tuning of the values at the single card level. Moreover, the 24 most exposed chambers will be replaced with new ones, equipped with high-quality (i.e. smoother surface) Bakelite laminates. The tests performed on the new FEE cards, used both in a test bench and on detectors, and on the new RPC chambers (with cosmic rays) are reported.

---

[*] Corresponding author.

# Contents

## 1. Introduction

ALICE [1] is one of the CERN LHC experiments: it is specifically designed to study ultrarelativistic nucleus−nucleus collisions and the Quark Gluon Plasma formed therein, together with proton−proton and proton−nucleus collisions which serve as a reference. ALICE is equipped with a Muon Spectrometer which measures the momentum of muons emitted in the forward rapidity region ($2.5<\eta<4$). The identification and selection of muons is performed by the MTR detector, based on Resistive Plate Chambers. MTR provides an estimate of the transverse momentum by measuring the track deflection with respect to an infinite momentum, i.e. straight track. This estimate allows selection of high momentum muons in order to increase the sample of events containing quarkonia and open heavy-flavour, as well as electroweak bosons.

     This system is running smoothly since the start of the LHC in 2010: however, after the two-year long shutdown starting in 2019, the ALICE detector will undergo a major upgrade in order to handle the increased LHC luminosity especially in Pb−Pb collisions. Since the selectivity of the rare probe triggers is low, ALICE decided to switch to continuous data-taking mode. Finally, the charge already accumulated by some RPCs is close to the lifetime of the detectors measured in dedicated ageing tests.

In order to cope with this forthcoming scenario, the Muon Trigger will be upgraded or several of its components will be replaced: in this paper, details of the upgrades are given.

## 2. The Muon Trigger

### 2.1 Description of the Muon Trigger system

The Muon Trigger is composed of 72 single-gap RPCs, arranged in 2 stations of 2 planes each and placed respectively at 16 and 17 m from the Interaction Point (IP), after a 120 cm thick iron wall (muon filter) located at the end of the tracking section of the spectrometer (Figure 1, left). Each plane area is about 5.5 x 6.5 m$^2$. The stations are arranged in projective geometry with respect to the Interaction Point: this means that detector dimensions, strip pitch and length increase from the first to the second station, according to the distance from the IP.

The RPCs are made in three different shapes (Long, Cut, Short) in order to accommodate the LHC beam pipe and its shielding (Figure 1, right). The total number of readout strips is about 21,000: the three different strip pitches (~1,2,4 cm wide, increasing with the distance from the beam pipe) allow one to keep the strip occupancy and the momentum resolution roughly constant all over the detection planes. The detector planes are mounted on mechanical frames that can be moved horizontally perpendicular to the beam line, in order to allow access to the chambers for replacement and/or maintenance purposes.

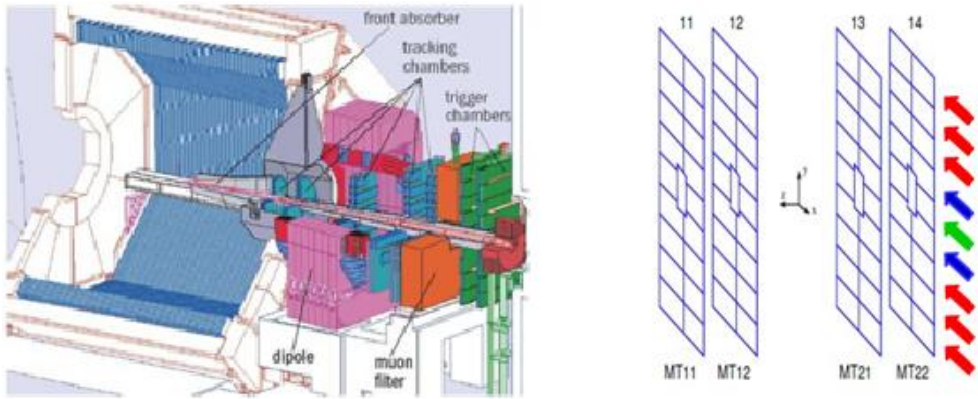

**Figure 1** - Left panel: the ALICE Muon Spectrometer, with the trigger stations in green color. Right panel: the layout of the trigger stations, with the arrows indicating the different shapes (red, blue and green indicate Long, Cut and Short detectors, respectively)

**2.2 The RPCs of the Muon Trigger**

The Muon Trigger RPCs [2] are single-gap detectors equipped with electrodes made of HPL (High Pressure Laminate) sheets. Both electrodes and gas gaps are 2 mm thick, the HPL resistivity is in the range between $3·10^9$ and $1·10^{10}$ Ω·cm. The gas gap dimensions range from about 210 x 70 cm$^2$ for the Short typology up to 280x80 cm$^2$ for the Long typology.
The gas gaps were manufactured by General Tecnica in Colli (FR, Italy) in 2004−2005 [3] and are operating in ALICE since 2010 (Run 1 2010−2013, Run 2 2015-2018). The working gas mixture consists of 89.7% $C_2H_2F_4$, 10% i-$C_4H_{10}$ and 0.3% $SF_6$.

The detector readout is made of segmented copper strip planes, manufactured at the INFN technological laboratory in Torino: analogic signals are discriminated by ADULT [4] Front End (FE) cards, with a threshold of 7 mV ("maxi-avalanche" operating mode [5]) without any pre-amplification stage. The FE cards issue LVDS digital signals to the Local and Regional cards that select muon tracks with a minimum transverse momentum larger than 0.5 GeV/$c$. The effective High Voltage (HV) applied is about 10400-10500 V at an atmospheric pressure of 970 mbar and a temperature of 20 °C. In these working conditions, the average charge per hit is about 100 pC and the maximum rate capability (measured in irradiation tests at the CERN Gamma Irradiation Facility) is about 100 Hz/cm$^2$ [6].

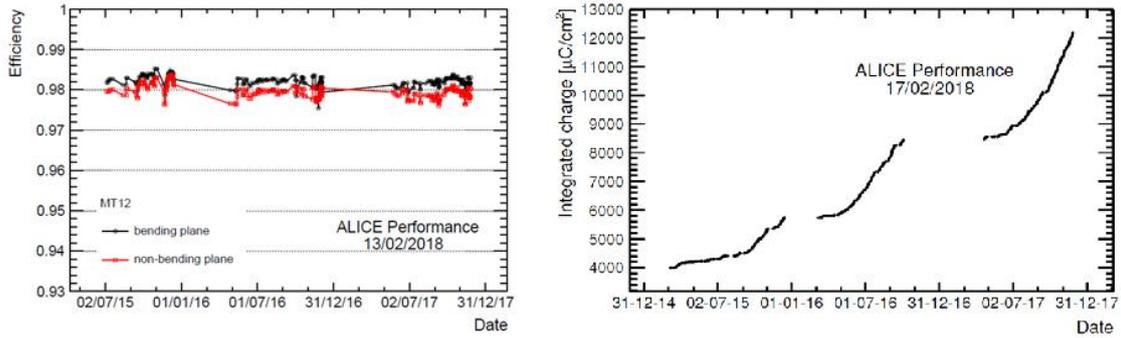

**Figure 2** - Left panel: average efficiency of the four RPC planes during LHC Run 2 (2015−2017 data). Right Panel: total integrated charge, averaged between the four detection planes,

The efficiency of the detector planes has been well within the requirements and stable during Run 2 (2015−2017 data), as shown in Figure 2 (left). The average integrated charge (Figure 2, right) is about 12 mC/cm$^2$, while the most exposed gaps reached about 30 mC/cm$^2$. The RPCs were ageing-tested up to 50 mC/cm$^2$, so this is their minimal certified lifetime.

## 3. The upgrade: from Muon TRigger to Muon IDentifier

From Run 3 (2021−onwards), the ALICE experiment will run in *continuous readout mode* (i.e. without trigger) and the Muon Trigger will therefore become a Muon Identifier. In addition to this, during LHC Run 3 and 4 the collision rate in ALICE will be raised and this will induce a large increase in hits on the RPCs: in Pb−Pb collisions, the maximum counting rate will rise from less than 10 Hz/cm$^2$ up to 90 Hz/cm$^2$, i.e. very close to the maximum rate capability of the detectors in maxi-avalanche mode. This increase will moreover induce an accelerated ageing of the gas gaps, which would reach the end of the expected lifetime well before the conclusion of Run 4, thus forcing a costly and time-consuming replacement of some gas gaps. In order to face these challenges a major upgrade is required, which is detailed in the following.

### 3.1 The FEERIC project: FE Electronics for the RPC detectors

It is well known that, in order to reduce the ageing and increase the rate capability, the charge delivered inside the gas gap has to be reduced. This goal can be achieved with the addition of a signal pre-amplification stage upstream of the front-end discriminators.

To this end, the new ASIC FEERIC (Front-End Electronics Rapid Integrated Circuit [7]), developed by the LPC Clermont-Ferrand group, includes a pre-amplification stage. Pre-series ASICs featured a response time dispersion smaller than 1.5 ns (to be compared with the 3 ns requirement) and a gain dispersion of about ±10%. 39 prototype FEERIC cards have been installed on one RPC in the ALICE cavern in February, 2015 and have shown very satisfactory performance and stability [8], allowing one to lower the working HV by several hundred volts (Figure 3). This lower HV allows a factor 4 reduction of the charge released in the RPC gas with respect to the previous ADULT cards, and we expect that the ageing will thus be reduced by a similar factor. The total ASIC production (5000 pcs.) was validated in 2 weeks in June 2016, and the yield was higher than 98%: the FE cards have been produced and were delivered in full by January, 2018.

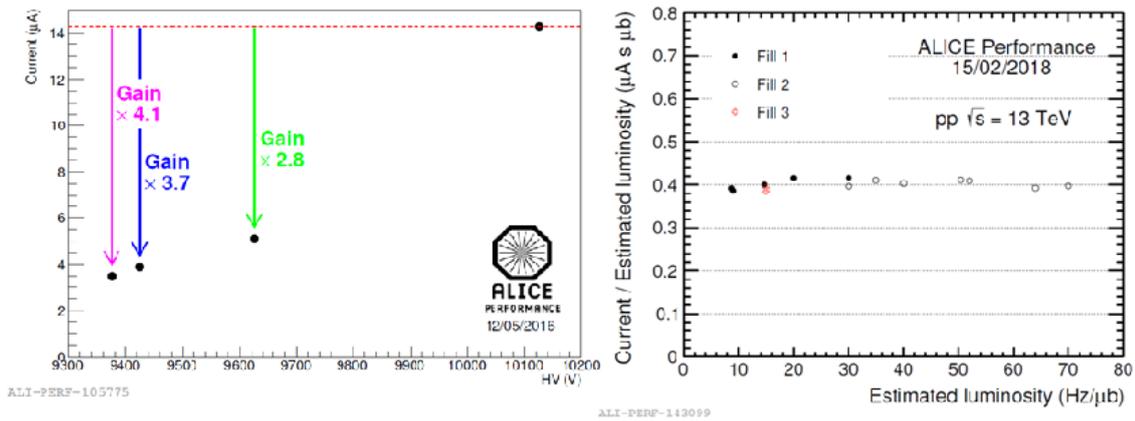

**Figure 3** Left: Current drawn at different HV working points (depending on different discrimination thresholds) by the RPC equipped with FEERIC. The black dot at the right of the plot represents the performance of the same RPC equipped with ADULT cards, without amplification stage. Right: Current drawn by the RPC normalized to the estimated luminosity as a function of luminosity itself.

### 3.2 Wireless FEERIC threshold distribution

The FEERIC card thresholds will be passed individually to each card by wireless connection and I2C chained per chamber side. This constitutes a large improvement with respect to the current configuration, in which the ADULT thresholds are set by means of an analog voltage distribution, allowing for one single threshold per RPC side. This new feature will allow one to minimize the RPC operating HV by tuning the thresholds according to the local RPC efficiency, which will therefore be more homogeneous. The choice for the wireless has fallen on the Zigbee technology, well suitable for required data bandwidth and speed. Its bandwidth is the 2.4 GHz/Radio Communication, based on Microcontroller Atmel SAMD21; the software is based on Arduino libraries (I2C, SD cards and Xbee). The checksum is performed to guarantee the correctness of data transmission.

This system is implemented on Xbee cards, designed by LPC group: 24 peripheral cards, placed close to detectors, plus 2 master cards (placed one on each side of the ALICE cavern) will be needed in total. The Xbee card for master and peripheral role is the same, as well as its firmware: the master/peripheral role is assigned by software configuration. By default, a board acts as peripheral.

Data contained on an SD Card are used for card initialization, while an EEPROM is used for storing/re-loading last used threshold values in case of a power cycle. During the 2017 year-end technical stop, one full line has been installed on the RPC already equipped with FEERIC cards in the cavern: the year 2018 will be devoted to long term stability tests.

### 3.3 Readout Electronics replacement

In order to switch to the continuous mode of data acquisition, the readout electronics (234 Local and 16 Regional FPGA-based cards, each one implemented on a dedicated PCB [Printed Circuit Board]) will be fully replaced. The new readout will be interfaced with the Common Readout Unit (CRU) developed for LHCb/ALICE. For R&D purposes, it was proposed to emulate, on a single PCB, 7 present Local cards connected by e-links@320 Mb/s to one Regional card implementing 1 GBT@3.2 Gb/s. This readout card prototype was ready by the end of 2015 and is fully operational since then. The pre-series validation, composed of individual Local and Regional cards as foreseen in the final project, is ongoing: the Production Readiness Review was held in April, 2018.

## 3.4 Test of the new RPCs

The adoption of the new FEERIC front-end discrimination will allow a sizeable extension of the lifetime of the RPCs. However, some RPCs have already accumulated a large integrated charge: in particular, some gas gaps already accumulated about 30 mC/cm$^2$ and will not be able to work properly until the end of Run 4. In order to take advantage of the prolonged and unrestricted access to the ALICE cavern during the long LHC shutdown of 2019−2020, the 24 most exposed RPCs (i.e., all the Short and Cut typologies, which are the ones closest to the beam pipe) will be replaced in 2019.

The new RPCs will be made with a new type of HPL electrodes, produced by the Puricelli factory in Costa Masnaga (MI, Italy), which feature a smoother surface with respect to the HPL used in the present detectors. The detectors are being manufactured at General Tecnica in Colli (FR, Italy), and are currently tested to check their performances. The tests are performed with cosmic rays in the INFN Torino laboratory and are similar to the ones performed in 2005 [3] on the presently installed RPCs. The tests will include:

- Detection of gas leaks;
- Check of the absence of ohmic leakage currents;
- Measurement of the efficiency vs. HV curve in cells of about 20 x 20 cm$^2$ to select the working HV;
- Measurement of the noise map of the detector;
- Measurement of the efficiency map at the working HV, with a granularity of about 2 x 2 cm$^2$.
- Measurement of electrodes resistivity, by filling the gas gap with Argon and measuring the current drawn when 2500 V are applied.

The efficiency will be measured by means of a cosmic ray tracking station, composed of 3 planes of 9 scintillators each plus 2 tracking RPCs placed in moving supports in order to cover the whole area of the RPCs under test. The RPCs are tested with a streamer mixture (50.5% Ar, 41.3% $C_2H_2F_4$, 7.2% i-$C_4H_{10}$, 1% $SF_6$) in order to be consistent with the previous tests in 2005. Analogic signals from the RPCs are processed by the ADULT front-end discriminators, with internal threshold of 80 mV. Two half RPCs can be tested at a time, and the time required to fully test two RPCs is about 5 days.

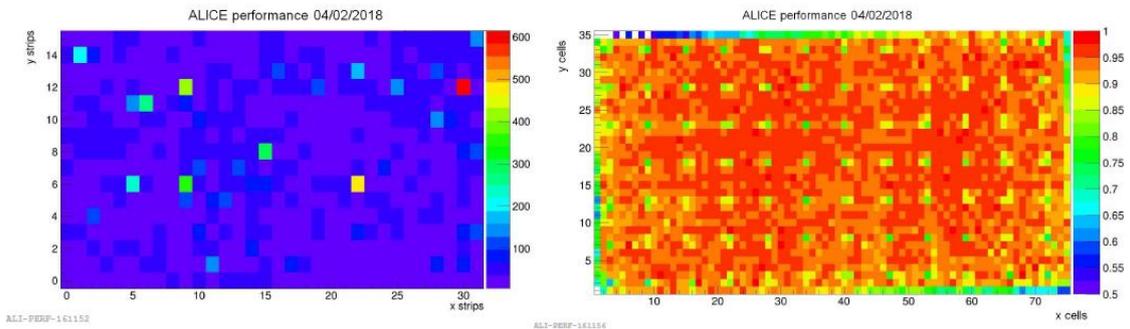

**Figure 4** – Left panel: noise map of one RPC half plane; the colour scale is in hundredths of Hz/cm$^2$, the red square corresponds to 6 Hz/cm$^2$. Right panel: efficiency map of one RPC half plane. The units on the axes correspond to the number of 2x2 cm$^2$ cells.

Triggering and DAQ are performed via three FPGA modules (CAEN V1495) connected to a PC. As of May, 2018 16 RPCs of the new production batch have been tested: the results of these first tests are however not fully satisfactory, because in some detectors the knee of the efficiency curve is reached at higher HV values (up to 8800 V, instead of customary 8300 V) and sometimes it is not uniform over their entire surface. This inhomogeneity could be due to a sub-optimal gluing of the gaps during manufacturing, with different amounts of glue on different spacers: this could induce variations on the gap thickness, which result in a variation of the effective electric field. In order to overcome this problem, the final production batch of the gas gaps will be done with a controlled amount of glue and under the supervision of the ALICE Torino group.

## 4. Conclusions

The ALICE Muon Trigger is running smoothly since the start of data-taking and its performance is fully satisfactory. The FEERIC project, which is part of the upgrade project, is in good shape: *in situ* tests proved fully satisfactory, the cards have been delivered and are ready for installation. The RPC tests are ongoing: some detectors show efficiency nonuniformities but this phenomenon does not affect the RPC global efficiency at the working point HV. Installation of the FEERIC and of new RPCs will take place on schedule starting in 2019.

## References

[1] ALICE collaboration, K. Aamodt et al. *The ALICE experiment at the CERN LHC*, Journal of Instrumentation, 2008, 3.08: S08002.

[2] R. Arnaldi et al., *A low resistivity RPC for the ALICE dimuon arm*, Nuclear Instruments and Methods in Physics Research Section A: Accelerators, Spectrometers, Detectors and Associated Equipment, Volume 451.2 (2000) 462-473.

[3] R. Arnaldi et al., *Overview on production and first results of the tests on the RPCs for the ALICE dimuon trigger,* Nuclear Physics B – Proceedings Supplement 158 (2006), 83-86.

[4] R Arnaldi et al*., A dual threshold technique to improve the time resolution of resistive plate chambers in streamer mode*, Nuclear Instruments and Methods in Physics Research Section A: Accelerators, Spectrometers, Detectors and Associated Equipment Volume 457, Issues 1–2 (2001), 117-125.

[5] R. Arnaldi et al*., Beam and aging tests with a highly-saturated avalanche gas mixture for the ALICE p-p data taking*, Nucl. Phys. B 158 (2006) 149.

[6] R. Arnaldi et al., *Ageing test of RPC for the Muon Trigger System for the ALICE experiment,* IEEE Symposium Conference Record Nuclear Science 2004., 2004, pp. 2072-2076 Vol. 4.

[7] S. Manen, P. Dupieux, B. Joly, F. Jouve and R. Vandaele, *FEERIC, a very-front-end ASIC for the ALICE muon trigger resistive plate chambers*, 2013 IEEE Nuclear Science Symposium and Medical Imaging Conference (2013 NSS/MIC), Seoul, 2013, pp. 1-4.

[8] ALICE collaboration, M. Marchisone, *Performance of a resistive plate chamber equipped with a new prototype of amplified front-end electronics* Journal of Instrumentation, 2016, 11 C06011.